\documentstyle[aps,prb]{revtex}
\newcommand{\be}{\begin{equation}}
\newcommand{\ee}{\end{equation}}
\newcommand{\bea}{\begin{eqnarray}}
\newcommand{\eea}{\end{eqnarray}}

\begin{document}

\title{Island nucleation in the presence of step edge barriers: \\
Theory and applications}
\author{Joachim Krug$^{1}$, Paolo Politi$^{1,2}$ and Thomas Michely$^{3}$}
\address{1. Fachbereich Physik, Universit\"at GH Essen,
D-45117 Essen, Germany \\
2. INFM, Unit\`a di Firenze, L.go E. Fermi 2, I-50125 Firenze, Italy \\
3. I. Physikalisches Institut, RWTH Aachen, D-52056 Aachen, Germany}
  \date{\today}

\maketitle

\begin{abstract}

We develop a theory of nucleation on top of two-dimensional islands 
bordered by steps with an additional energy
barrier $\Delta E_S$ for descending atoms. The theory is based
on the concept of the {\em residence time} of an adatom on the island,
and yields an expression for the nucleation rate which becomes exact
in the limit of strong step edge barriers. This expression differs
qualitatively and quantitatively from that obtained using the
conventional rate equation approach to nucleation [J. Tersoff {\em et al.},
Phys. Rev. Lett. {\bf 72}, 266 (1994)]. We argue that rate equation theory
fails because nucleation is dominated by the rare instances when two
atoms are present on the island simultaneously. The theory is applied to
two distinct problems: The onset of second layer nucleation in 
submonolayer growth, and the distribution of the sizes of top terraces
of multilayer mounds under conditions of strong step edge barriers. 
Application to homoepitaxial growth on Pt(111) yields the estimate
$\Delta E_S \geq  0.33$ eV for the 
additional energy barrier at CO-decorated steps.


\end{abstract}

\section{Introduction}
\label{Intro}
On many metal surfaces an adatom descending across a
step edge encounters an additional energy barrier
$\Delta E_S$ compared to the diffusion barrier $E_D$  on an
atomically flat terrace \cite{ehrlich}. This {\em step edge barrier}
controls the rate of interlayer mass transport and therefore
has a decisive influence on the morphology of multilayer films 
\cite{comsa,villain,lagally}. 

A number of papers have addressed
the question of how to determine $\Delta E_S$ experimentally
\cite{tersoff,meyer,pavel,bromann,markov,roos97,roos98}. 
Most approaches start from the observation that the step edge
barrier increases the residence time of adatoms on top of an
island, and therefore promotes the nucleation of the next layer.
Tersoff, Denier van der Gon and Tromp \cite{tersoff} (TDT) 
presented a quantitative analysis of this effect, which is
based on the conventional estimate \cite{venables}
\begin{equation}
\label{rateeq}
\omega \sim \nu n^{i^\ast + 1}
\end{equation}
for the nucleation rate $\omega$ in terms of the in-layer hopping rate
$\nu$, the adatom density $n$ and the size $i^\ast$ of the
largest unstable cluster. Equation (\ref{rateeq}) arises from
rate equations for spatially averaged island and adatom densities,
and its applicability to the confined geometry on top of an island
is not obvious. Other treatments invoke the concept of a 
critical adatom density for nucleation, which is defined as the
density at which nucleation would take place on the unbounded
terrace \cite{meyer,pavel}. Again, given the rather different conditions
on top of an island, this approach seems hard to justify
\cite{markov} (see Section \ref{Critical_density}). 

In this paper we present a detailed microsopic analysis
of nucleation in the presence of strong step edge barriers,
which takes into account the large fluctuations of the adatom
population on top of the island. Our expression for the nucleation
rate, to be derived in Section \ref{Nucleation}, differs
qualitatively and quantitatively from that obtained by TDT. 
The rate equation treatment significantly overestimates the nucleation
probability, and therefore the analysis of 
experimental data based on the TDT theory generally gives values
for $\Delta E_S$ which are too small. 
To illustrate our point, we
apply our approach to two different experimental situations.
In Section \ref{Secondlayer} we determine the critical radius for
second layer nucleation in submonolayer deposition, and 
use it to reanalyse
the experiments of Bromann et al.\cite{bromann} for Ag(111),
and of Kalff et al.\cite{Kalff98} for CO-contaminated Pt(111). In
Section \ref{Wedding} we develop a simple analytic theory for the
size of the top terrace of pyramidal multilayer mounds
(``wedding cakes'' \cite{jk97,politi97}), and apply
it to the case of Pt(111) \cite{kalff99}. For CO-decorated
steps on Pt(111) the two approaches yield mutually consistent
estimates for $\Delta E_S$. Some of the results presented
in Section \ref{Secondlayer} have been independently obtained by
Rottler and Maass\cite{maass99}, who also provide numerical confirmation
by Monte Carlo simulations
for $i^\ast = 1$.

\section{Nucleation rate on top of an island}

\subsection{Irreversible aggregation with strong barriers}
\label{Nucleation}

We consider a compact two-dimensional island of area $A$ and 
perimeter $L$. Both $A$ and $L$ are dimensionless quantities 
measured in terms of the number of lattice sites on the island ($A$)
and the number of edge sites ($L$), respectively. 
For large $L$ they are related through
\begin{equation}
\label{alpha}
A = \alpha L^2
\end{equation}
where the coefficient $\alpha$ depends on the island shape; for example,
$\alpha = 1/16$ for square islands, $\alpha = 1/12$ for regular hexagons on
a triangular lattice and $\alpha = 1/4 \pi$ for a circular island
in a continuum approximation. Atoms are
deposited at rate $F$, diffuse on the island (and on the underlying terrace)
at rate $\nu = \nu_0 \exp[- E_D/k_B T]$, 
and descend from edge sites at rate $\nu' = 
\nu_0' \exp[-E_S/k_B T]$. The additional energy
barrier at the step edge is then $\Delta E_S = 
E_S - E_D$. Since the actual energy landscape on real islands
may well be more complicated \cite{ehrlich97,kyuno98},
$\nu'$ should be regarded as an {\em effective} interlayer
transport rate. 

The problem involves three time scales: The
traversal time\cite{tang}
\begin{equation}
\label{tautr}
\tau_{\rm tr} \approx  A/\nu
\end{equation} 
required for an atom to visit
all sites of the island, the mean time interval
\begin{equation}
\label{Deltat}
\Delta t = (F A)^{-1}
\end{equation}
between subsequent deposition events on the island, 
and the mean residence time $\tau$ 
that a single adatom spends on the island. 
The latter is quite generally given in terms of the mean
adatom density $\bar n$ and the flux $F$ as
\begin{equation}
\label{taugen}
\tau = {\bar n}/F.
\end{equation}
To arrive at this relation, consider 
the deposition of noninteracting atoms onto the island
during some long time interval $T$.
A total of $T/\Delta t$ atoms arrive on the island, each of
which spends an average time $\tau$. The total time spent by all 
the atoms is $T \tau /\Delta t$, and hence the mean number of atoms
simultaneously present on the island is 
$\tau/\Delta t =  \bar n A $,
from which (\ref{taugen}) follows. 

Here we focus on the case of {\em strong} step edge barriers
(case 2 of TDT), in the sense that
the Ehrlich-Schwoebel length \cite{pv}
\begin{equation}
\label{les}
\ell_{\rm ES} = (\nu/\nu' -1) \approx \nu/\nu'
\end{equation}
is much larger than the island size. 
In this limit the adatom density on the island becomes spatially uniform
\cite{tersoff}, and its value is determined simply by the balance between
the number of atoms $F A$ deposited on the island, and the
number of atoms $L {\bar n} \nu'$ leaving it per unit time 
\cite{lagally,michely}. Thus $\bar n = F A/L \nu' = \alpha F L/\nu'$
and 
\begin{equation}
\label{tau}
\tau = \frac{\alpha L}{\nu'}. 
\end{equation}
It follows that $\tau/\tau_{\rm tr} \approx \ell_{\rm ES}/L$, and
therefore the strong barrier condition is equivalent to 
$\tau/\tau_{\rm tr} \gg 1$. 

In this section we further assume that the critical island size takes
its minimal value $i^\ast = 1$, i.e. dimers are stable and immobile. Then,
as soon as two adatoms are present on the island,
nucleation occurs after a time of the order of $\tau_{\rm tr} \ll 
\tau$;
on the scale of the residence time nucleation is an instantaneous event. 
We conclude, therefore, that the nucleation probability 
$p_{\rm nuc}$ per deposited
adatom is equal to the probability $p_2$ that two adatoms are present
simultaneously on the island. To compute $p_2$, suppose the first 
adatom has arrived on the island at time $t=0$, and denote by $t_1$ and
$t_2$ the departure time of the first adatom and the arrival time of the
second one. Since deposition is a Poisson process, $t_2$ is an exponential
random variable with mean $\Delta t$.
In the strong barrier limit also the 
distribution of residence times $t_1$ decays exponentially\cite{harris} 
for times large compared to $\tau_{\rm tr}$. 
Thus we have
\begin{equation}
\label{p2}
p_{\rm nuc} = p_2 = {\rm Prob}[t_1 > t_2] = 
\frac{1}{\tau \Delta t} \int_0^\infty dt_1 \; e^{-t_1/\tau} 
\int_0^{t_1} dt_2 \; e^{-t_2/\Delta t} = \frac{\tau}{\tau + \Delta t}.
\end{equation}
If $\Delta t \ll \tau$ 
we find $p_2 \approx 1$, i.e. the first adatom
deposited on the island already forms a nucleus. In this limit the nucleation
rate becomes independent of $\nu'$, which means that the step edge barrier is,
at least as far as second layer nucleation is concerned, infinite. 
We will return to this case at the end of Section \ref{Critical}.
In the following we will mostly be interested in
the opposite limit $\Delta t \gg \tau$, where
$p_2 \approx \tau/\Delta t \ll 1$. A general expression for the probability
$p_{n+1}$ to find $n+1$ adatoms simultaneously on the island is derived
in Appendix \ref{pn+1}. 

The expression for $p_{\rm nuc}$ can also be found by using an
argument first developed in one dimension by Elkinani and Villain~\cite{EV}
(see Ref.~25 for the two-dimensional case).
According to this approach, 
the nucleation probability per deposited atom is simply
the number of distinct sites visited by each adatom (equal to
$A$) times the probability that a site is occupied (given by the
adatom density $\bar n$). Together with (\ref{taugen}) this implies
$p_{\rm nuc} = A F \tau = \tau/\Delta t$ as above. 

To obtain the nucleation rate $\omega$, defined as
the number of nucleation events taking place on
the island per unit time, we multiply the nucleation probability $p_{\rm nuc}$
by the number of atoms arriving on the island per unit time, and obtain,
in the relevant case $\Delta t \gg \tau$,
\begin{equation}
\label{omega}
\omega = F A \frac{\tau}{\Delta t} = \frac{\alpha^3 F^2 L^5}{\nu'}.
\end{equation}
This is to be compared to the expression
\begin{equation}
\label{tersoff}
\omega = \gamma \frac{\alpha^3 F^2 L^4 \nu}{4 \nu'^2}
\end{equation}
derived from the approach of TDT 
($\gamma$ denotes a capture number
of order unity). Equation (\ref{tersoff}) exceeds (\ref{omega}) 
by a factor $(\nu/\nu') L^{-1} = \ell_{\rm ES}/L \gg 1$. 
To explain this discrepancy we note that the
mean number of adatoms on the island is
much less than unity in the regime of interest.
Most of the time the island is empty, and sometimes it is occupied by a 
single adatom. 
For example, consider a Pt island on
Pt(111) without a second layer nucleus,
but at a size, 
at which about half of the island population has already
formed such a nucleus (see Section \ref{Pt(111)islands}). 
Even for this island the probability to find an atom on top
is only about $10^{-2}$, as is the probability
for an atom to become deposited on an island already populated by an
atom. For this reason we refer to our approach as the
{\em lonely adatom model} (LAM). The strong fluctuations in the
occupancy of the island imply that
the replacement of the actual adatom density by its time average $\bar n$,
which is inherent in (\ref{rateeq}), 
cannot be justified, and must be replaced by the more
detailed statistical analysis provided above.

\subsection{Intermediate and weak barriers}
\label{Weak}

To some extent the considerations of the preceding section can be
generalized to the case when the strong barrier condition $\ell_{\rm ES}
\gg L$ is no longer satisfied. The mean residence time is still
given by (\ref{taugen}), and the mean adatom density $\bar n$
can be obtained by solving the stationary diffusion equation with
the appropriate boundary conditions \cite{tersoff,pavel,review}. 
For the case of a 
circular island of radius $r = L/2 \pi$
this yields 
\be
\label{tauweak}
\tau = \frac{1}{2 \nu} \left(r^2 +  
\frac{\nu r}{\nu'} \right).
\ee
However now the probability $p_2$ provides only 
an upper bound on the nucleation
probability $p_{\rm nuc}$, since one of the two adatoms may escape from
the island before the two meet. 
Also in the evaluation of $p_2$
in eq.(\ref{p2}) the probability distribution $P_{\rm res}(t_1)$
of residence times has to be used, which is a complicated non-exponential
function \cite{harris}. The calculation simplifies 
by noting that for weak barriers surely $\tau \ll \Delta t$;
then the exponential distribution of interarrival times can be
expanded and one obtains
\be
\label{p2weak}
p_2 = 
\int_0^\infty dt_1 \; P_{\rm res}(t_1)
(1 - e^{-t_1/\Delta t}) \approx \frac{1}{\Delta t}
\int_0^\infty dt_1 \; t_1 P_{\rm res}(t_1) = \frac{\tau}{\Delta t}.
\ee
We conclude that an upper bound on the nucleation rate for 
a circular island is given by the expression
\be
\label{bound}
\omega_> = \frac{\pi^2 F^2 r^4}{2 \nu} \left(r^2 +  
\frac{\nu r}{\nu'} \right).
\ee 
If the barriers are weak, $r \gg \ell_{\rm ES}$, we find
$\omega_> = \pi^2 F^2 r^6/2\nu$, which agrees (up to a numerical
prefactor) with Eq.~(4a) of TDT. 
This implies that the rate equation ansatz (\ref{rateeq}) 
yields the correct result only 
when the barriers can be neglected.

\subsection{Reversible aggregation}
\label{I2}

We return to the case of strong step edge barriers and
analyze reversible aggregation with stable trimers and unstable dimers
($i^\ast = 2$).
The dimer dissociation time $\tau_{\rm dis}$
introduces a further time scale into the problem. Suppose that two
adatoms are present on the island; this is true with probability
$p_2 = \tau/\Delta t$. 
Let us assume for simplicity that one atom is fixed and the other
diffusing. If the two adatoms stay on the island a time $t$,
this means that the diffusing
atom has failed to descend a number of times of order
$ j = t/(\tau_{\rm tr} + \tau_{\rm dis})$, which is true with a probability
of the order of order $\exp (-j\tau_{\rm tr}/\tau)$. 
Therefore the effective residence time $\tau'$ 
of the adatom pair can be estimated as 
$\tau' \sim \tau (\tau_{\rm tr} + \tau_{\rm dis})/\tau_{\rm tr}$.
In the limit $\tau_{\rm tr} \gg \tau_{\rm dis}$, $\tau' \sim \tau$, and
in the limit $\tau_{\rm tr} \ll \tau_{\rm dis}$, $\tau' \sim 
\tau (\tau_{\rm dis}/\tau_{\rm tr})$. The lifetime of the pair
is increased compared to the single adatom lifetime if
$\tau_{\rm tr} \ll \tau_{\rm dis}$. 

Consider first the case $\tau_{\rm dis} \ll \tau_{\rm tr}$, when
the total residence time of the two adatoms is still of the order of
$\tau$, and a dimer is present a fraction $q \equiv
\tau_{\rm dis}/\tau_{\rm tr}$
of that time. The probability for a third atom to deposit while the
two atoms are on the island is $\tau/\Delta t$ as before. The third
atom traverses the island 
of the order of $m = \tau/\tau_{\rm tr}$
times, and each time
the probability that it encounters a dimer is $q$. We therefore
have to distinguish the cases $m q \ll 1$ and $m q \gg 1$. In the 
first case the probability that a stable trimer forms is 
$m q \sim \tau \tau_{\rm dis}/\tau_{\rm tr}^2$, 
and altogether the nucleation
probability per atom is of the order of
\begin{equation}
\label{caseI}
p_{\rm nuc} \sim 
\left(\frac{\tau}{\Delta t} \right)^2 
\frac{\tau \tau_{\rm dis}}{\tau_{\rm tr}^2}
\sim \frac{F^2 L^3 \nu^2 \tau_{\rm dis}}{\nu'^3} 
\;\;\;\;\;\;\; 
{\rm for} \;\;\; \tau_{\rm dis} \ll \tau_{\rm tr}^2/\tau
\;\;\;
({\rm regime \;\; I}).
\end{equation}
On the other hand, if $m q \gg 1$ the third atom is certain to 
encounter a dimer, and 
\begin{equation}
\label{caseII}
p_{\rm nuc} \sim 
\left(\frac{\tau}{\Delta t} \right)^2 
\sim \frac{F^2 L^6}{\nu'^2} 
\;\;\;\;\;\;\; 
{\rm for} \;\;\; 
\tau_{\rm tr}^2/\tau \ll \tau_{\rm dis} \ll \tau_{\rm tr}
\;\;\;
({\rm regime \;\; II}).
\end{equation}
In fact a somewhat more precise statement can be made. In regime II
the nucleation probability is equal to the probability $p_3$ of finding
three adatoms simultaneously on the island. The calculation in Appendix 
\ref{pn+1}
shows that, for $\tau \ll \Delta t$, $p_3 = (1/2) (\tau/\Delta t)^2$, 
and therefore the prefactor in (\ref{caseII}) is $\alpha^4/2$. 

Next consider the case $\tau_{\rm dis} \gg \tau_{\rm tr}$,
when the effective residence time of the first two adatoms is 
$\tau' \sim \tau (\tau_{\rm dis}/\tau_{\rm tr}) \gg \tau$. 
If a third
atom is deposited onto the island it will find the dimer with
a probability close to unity. A third atom will deposit during
time $\tau'$ with probability $\tau'/\Delta t$ if 
$\tau' < \Delta t$ and with probability one if $\tau' > \Delta t$.
This implies two further scaling regimes:
\begin{equation}
\label{caseIII}
p_{\rm nuc} \sim 
\frac{\tau}{\Delta t} \cdot \frac{\tau'}{\Delta t} 
\sim \frac{F^2 L^4 \nu \tau_{\rm dis}}{\nu'^2} 
\;\;\;\;\;\;\; 
{\rm for} \;\;\; 
\tau_{\rm tr} \ll \tau_{\rm dis} \ll \Delta t \cdot (\tau_{\rm tr}/
\tau) \;\;\;
({\rm regime \;\; III}).
\end{equation}
\begin{equation}
\label{caseIV}
p_{\rm nuc} \sim 
\frac{\tau}{\Delta t} 
\sim \frac{F L^3}{\nu'} 
\;\;\;\;\;\;\; 
{\rm for} \;\;\;
\tau_{\rm dis} \gg
\Delta t \cdot (\tau_{\rm tr}/\tau) 
\;\;\;
({\rm regime \;\; IV}).
\end{equation}
In the last regime the dimer lifetime is so long that the situation 
effectively reduces to the case $i^\ast = 1$.

The smallest value that $\tau_{\rm dis}$ can take is the inverse
in-layer hopping frequency $1/\nu$. Then $\tau_{\rm dis}/\tau_{\rm tr}
\sim L^{-2} \ll 1$ always, but depending on the strength of the 
step edge barrier both regimes I and II can be realized, corresponding
to $L^3 \gg \ell_{\rm ES}$ (regime I) and 
$L^3 \ll \ell_{\rm ES}$ (regime II), respectively. 
Setting $\tau_{\rm dis} = 1/\nu$ in 
regime I (Eq. (\ref{caseI})) yields
\be
\label{caseTDT}
p_{\rm nuc} \sim \frac{F^2 L^3 \nu}{\nu'^3} \;\;\;\;\;\;\; 
({\rm regime \;\; I,} \;\; \tau_{\rm dis} = 1/\nu)
\ee
which agrees with the expression derived by TDT
for $i^\ast = 2$.
This indicates, as suggested by Rottler and Maass \cite{maass99}, 
that the rate equation approach describes a situation in which
the probability for an encounter between the atoms forming the
stable nucleus (the quantity $m q$ in our derivation) is small
compared to unity.

The complexity of the case $i^\ast = 2$ 
illustrates the difficulty of extending the present analysis
to larger values of $i^\ast$. The generalization seems straightforward
only in regime II, where the barriers are sufficiently strong to 
guarantee the formation of a stable nucleus once $i^\ast + 1$ atoms
are on the island simultaneously (but still so weak that
$\tau \ll \Delta t$). Then $p_{\rm nuc} = p_{i^\ast + 1}$, and
using Eq.(\ref{pn2}) of the Appendix we obtain
\be
\label{caseIIgen}
p_{\rm nuc} = \frac{1}{i^\ast !} \left( 
\frac{\tau}{\Delta t} \right)^{i^\ast} = 
\frac{1}{i^\ast !} \left( 
\frac{\alpha^2 L^3 F}{\nu'} \right)^{i^\ast} \;\;\;\;\;\;\;
({\rm regime \;\; II, general} \; i^\ast).
\ee

\section{Second layer nucleation}
\label{Secondlayer}

\subsection{The critical island size}
\label{Critical}

In this section we consider a population of submonolayer islands
and ask for the fraction $f$ of islands on which a second layer
has nucleated. Once the nucleation rate $\omega(L)$ per island is
known as a function of island size $L$, 
this fraction is given by \cite{tersoff}
\begin{equation}
\label{f}
f = 1 - \exp[- \int dt \; \omega(L(t))],
\end{equation}
where the integration extends over the deposition time, and the
functional dependence $L(t)$ depends on the growth situation. 
For example, if first layer islands of density $N$ nucleate at time
$t = 0$, and no further nucleation occurs
at later times, then the island size increases with coverage
$\theta = F t$ as $L = \sqrt{\theta/\alpha N}$. 
Assuming a size dependence of the nucleation rate of the general form
\be
\label{omegagen}
\omega(L) = F \Omega L^k,
\ee
where $\Omega$ is a constant independent of $L$, one finds
\begin{equation}
\label{f(L)}
f = 1 - \exp[-(L/L_c)^{k+2}]
\end{equation}
where the critical island size $L_c$ is given by
\be
\label{Lc1}
L_c = \left(\frac{k+2}{2 \alpha N \Omega} \right)^{1/(k+2)}.
\ee
In particular, using our expression (\ref{omega}) for $i^\ast = 1$
we obtain 
\begin{equation}
\label{Lc}
L_c =  
\left( \frac{7}{2 \alpha^4} \cdot \frac{\nu'}{F N} \right)^{1/7}.
\end{equation}
This agrees with the scaling law derived by Rottler and Maass 
\cite{maass99} and specifies the prefactor. 

The different regimes for $i^\ast = 2$ discussed in Section
\ref{I2} can be treated in the same way; the results are summarized
in Appendix \ref{LcI2}. 
Inspection shows that our regime II corresponds to regime
II of Rottler and Maass\cite{maass99}, 
while our regime I corresponds to their
regime III (they assume that the dimers are maximally unstable
in the sense that $\tau_{\rm dis} = 1/\nu$). 

A critical island size for second layer nucleation can also be defined
when the barriers are so strong (or the flux so large) that
the condition $\tau \ll \Delta t$ is no longer fulfilled. Then
the nucleation probability per atom becomes of order unity, and
the nucleation rate on top of the island is $\omega \approx
F A$. Repeating the above analysis for this
case, one finds that 
\begin{equation}
\label{rcbig}
f = 1 - \exp[- \theta^2/2 N] = 1 - \exp[-(L/L_{\infty})^4]
\end{equation}
with $L_{\infty} = (2/\alpha^2 N)^{1/4}$. 
The critical island size is of the order of the square root of the
island spacing, independent of 
$\nu'$ and $i^\ast$. This corresponds to regime I of Rottler and
Maass \cite{maass99}.

\subsection{Conditions for layer-by-layer growth and surfactant action}
\label{Surfactants}

According to classical nucleation theory \cite{venables} the 
first layer island density depends on flux and in-layer mobility according to
the scaling law
\be
\label{N}
N \sim (F/\nu)^{i^\ast/(i^\ast + 2)}.
\ee
Provided the in-layer mobility on the island is the same as on the
terrace, for $i^\ast = 1$ the LAM expression (\ref{Lc}) can therefore
be written as
\be
\label{LcLAM}
L_c \sim \frac{\nu^{1/21} \nu'^{1/7}}{F^{4/21}} \;\;\;\;\;\;\; ({\rm LAM}),
\ee
while the corresponding expression of the TDT theory \cite{tersoff} reads
\be
\label{LcTDT}
L_c \sim \frac{\nu'^{1/3}}{\nu^{1/9} F^{2/9}} \;\;\;\;\;\;\; ({\rm TDT}).
\ee
Surprisingly, a decrease of the in-layer mobility $\nu$ is seen to
(weakly) {\em decrease} the critical island size according to the LAM,
but {\em increases} it according to the TDT theory. 

A decrease of the in-layer mobility is believed to play an important
role in the ability of certain adsorbates (such as 
Sb on Ag(111)) to act as {\em surfactants}, in the sense of promoting
layer-by-layer growth \cite{meyer,vrijmoeth94,scheffler95}. 
The transition from three-dimensional to layer-by-layer growth
occurs \cite{tersoff} when the critical island size becomes comparable
to the distance $N^{-1/2}$ between first layer islands. Equating 
(\ref{LcLAM}) and (\ref{LcTDT}) to $(\nu/F)^{1/6}$ one finds in both
cases $L_c \sim \ell_{\rm ES} \sim \nu/\nu'$, which shows that
the condition for the onset of layer-by-layer growth does not depend
on the details of the second layer nucleation mechanism, at least
as long as $\ell_{\rm ES}$ is the only additional length scale in the problem;
in the case $i^\ast = 2$ the dimensionless number $\tau_{\rm dis} \nu$
introduces another scale, and the situation becomes more complicated. 

\subsection{Adatom density at second layer nucleation}
\label{Critical_density}

It was mentioned in Section \ref{Intro} that, in addition to the rate
equation approach \cite{tersoff,markov}, the problem of second layer
nucleation has been treated using the concept of a critical density
for nucleation \cite{meyer,pavel}. To see whether such a concept
is meaningful in the present context, we can estimate the mean adatom
density $n_c$ on top of a first layer island at 
the time of second layer nucleation.
Using (\ref{taugen}) and (\ref{tau}) we obtain $n_c = F \alpha L_c/\nu'$,
and inserting the expression (\ref{Lc}) yields 
\be
\label{nc}
n_c \sim (F/\nu')^{6/7} N^{-1/7} \sim L_c^{-6} N^{-1}. 
\ee
This is to be compared to the adatom density $n_{\rm sub}$ 
on the substrate at the time of first layer nucleation, which 
is of the order of \cite{meyer,pavel} $(F/\nu) N^{-1} \sim N^2$ for
$i^\ast = 1$. The comparison
shows that $n_c \gg n_{\rm sub}$ provided $L_c$ is small compared to 
the mean island spacing, which is equivalent to the condition of
strong step edge barriers (see Section \ref{Surfactants}). Thus in the
regime of interest, the assumption $n_c \approx n_{\rm sub}$ 
is not satisfied, and consequently the estimates of $L_c$ obtained
in Refs.~6 and 7 are not quantitatively accurate.

\subsection{Nucleation on predeposited islands: The case of Ag(111)}

A detailed experimental study of second layer nucleation 
on Ag(111) was performed by Bromann et al.\cite{bromann}. 
They prepared arrays of 
approximately circular islands of uniform initial radius $r_0$
through Ostwald ripening at high temperatures.
Given the initial coverage $\theta_0$, the island density is 
then $N = \theta_0/\pi r_0^2$. Subsequently a second dose of coverage
$\Delta \theta$ was deposited, and the fraction $f$ of islands with
second layer nuclei was measured as a function of $r_0$. 
Independent evidence shows that the critical nucleus size
is $i^\ast = 1$ under the experimental growth conditions. 
Provided further nucleation during the second dose can be 
neglected \cite{bromann},
the island radius increases with coverage according
to $r = r_0 \sqrt{1 + \theta/\theta_0}$. Inserting this into 
(\ref{omega}) and (\ref{f}) and using that $\theta_0 = \Delta 
\theta$ in the experiment, we find
\begin{equation}
\label{fbromann}
f = 1 - \exp[-(r_0/{\tilde r_c})^5]
\end{equation}
with 
\begin{equation}
\label{rctilde}
\tilde r_c = \left( \frac{7}{\pi^2 (2^{7/2} - 1)} \cdot
\frac{\nu'}{F \Delta \theta} \right)^{1/5}.
\end{equation}

The step edge barrier can be extracted directly by comparing the 
critical radii at two different temperatures $T_1$ and $T_2$,
since according to (\ref{rctilde})
\begin{equation}
\label{twoT}
\tilde r_c(T_1)/\tilde r_c(T_2) = 
\exp[E_S(T_2^{-1} - T_1^{-1})/ 5 k_B].
\end{equation}
From Figure 2 of Ref.~8 we estimate that 
$\tilde r_c \approx 28 {\rm \AA}$ at 120 K and 
$\tilde r_c \approx 52 {\rm \AA}$ at 130 K. This yields
$E_S \approx$ 0.42 eV, leading to an additional step
edge barrier $\Delta E_S \approx 0.32$ eV. 
The theory of TDT predicts instead
that $\tilde r_c \sim (\nu'^2/\nu)^{1/4}$, which implies 
\begin{equation}
\label{twoT2}
\tilde r_c(T_1)/\tilde r_c(T_2) = 
\exp[(2E_S - E_D)(T_2^{-1} - T_1^{-1})/ 4 k_B]
\end{equation}
and yields the estimate $\Delta E_S \approx 0.12$ eV reported 
by Bromann et al.\cite{bromann}.

The dramatic discrepancy between the two estimates for $\Delta E_S$
makes it important\cite{roos98} 
to examine also the prefactor $\nu_0'$ in the
expression for $\nu'$. Using (\ref{rctilde}) with
$E_S = 0.42$ eV yields
$\nu_0' \approx 8 \times 10^{19} \; {\rm s}^{-1}$, which seems much
too large to be physically reasonable. Smaller values of $\nu_0'$
can be obtained if a (slight) temperature dependence of the
barrier is permitted. Since the step edge barrier of interest
here in fact constitutes an average over the (temperature-dependent)
step structure, such a dependence could be expected due thermal
roughening of the step \cite{kalff99}. For example, if we
demand that the prefactor be the same as for in-layer 
transport\cite{bromann},
$\nu_0' = \nu_0 = 2 \times 10^{11} \; {\rm s}^{-1}$,
then (\ref{rctilde}) yields $E_S \approx 0.21$ eV at $T = 120$ K and
$E_S \approx 0.20$ eV at $T = 130$ K, somewhat {\em smaller} than
the estimate $E_S \approx 0.22$ eV
of Bromann et al. \cite{bromann}, while taking their value 
$\nu_0' = 10^{13} \; {\rm s}^{-1}$ for the prefactor, 
yields the larger barrier
$E_S = 0.26$ eV ($T = 120$ K) and $E_S = 0.24$ eV 
($T = 130$ K), respectively. The effective barrier decreases with
increasing temperature, which is consistent with the step roughening
picture. Clearly a more definite statement about the values of 
$E_S$ and $\nu_0'$ on the Ag(111) surface requires a detailed
reanalysis of the experimental data, and perhaps also additional
measurements.

\subsection{Application to Pt(111) islands with decorated steps}
\label{Pt(111)islands}

In this section we use the LAM to analyze
the dependence of $\nu'$ and
$\Delta E_S$ on the CO partial pressure $p_{\rm CO}$ in Pt homoepitaxy on
Pt(111). CO exposure during
growth leads to step decoration, which impairs the
hopping of adatoms over the step edge and therefore
increases $\Delta E_S$ (for details see Ref.~13).

Rather than following the evolution of the fraction $f$ of islands
with second layer nuclei with coverage or island size, we employed
a method which requires only a single growth experiment
within the appropriate coverage range, i.e. a growth experiment
which creates islands partly with and partly without second layer
nuclei. The practical analysis is performed by determination of the
number of edge sites $L$ of the largest islands without a second
layer island on top and of the smallest islands with a second layer
one on top for a statistically significant number of topographs. The
averages of the two sets of values vary typically by less than 10
\%. The mean of the two averages 
then gives an estimate of the island size $L_{1/2}$ 
defined by $f(L_{1/2}) = 1/2$: at this island size 
it is equally likely to find an island with or without a second
layer nucleus. Using Eqs.(\ref{omega},\ref{f(L)},\ref{Lc})
for $i^\ast = 1$, one obtains the expression  
\begin{equation}
\label{method2}
\nu' = \frac{2}{7 \ln 2} \alpha^4 F N L_{1/2}^7
\end{equation}
for $\nu'$ in terms of $L_{1/2}$. 

In order to translate $\nu'$ into the additional step edge barrier
$\Delta E_S$, it is necessary to make 
assumptions on the adatom migration energy $E_D$ and the attempt
frequency $\nu_0'$ for hopping over the step edge. For the case of
Pt adatoms on Pt(111), $E_D$ was found to be $0.26$ eV in two
independent experimental studies\cite{Bott,Ehrlich98}, and the prefactors 
agreed within a factor of two with an average of $\nu_0 = 8 
\times 10^{12} {\rm s}^{-1}$.
An estimate of $\nu_0'$ for the clean case can be extracted
from a Field Ion Microscopy study, in which the adatom residence time
was directly measured \cite{kyuno98}. Using the relation (\ref{tau})
for circular islands, the prefactor obtained in Ref.~19 
translates into $\nu_0' \approx 7 \times 10^{11} {\rm s}^{-1}$, somewhat
smaller than $\nu_0$. However, since nothing is known about 
the effect of CO decoration on $\nu_0'$, in the following we take 
$\nu_0'$ to be identical to $\nu_0$ for simplicity.
The growth experiments\cite{Kalff98}
were carried out at 400 K with a deposition rate $F = 5 \times 10^{-3}$
ML/s.

The open circles in Fig. \ref{Fig-CO} exhibit 
the critical size $L_{1/2}$ as a function of
the applied CO partial pressure during growth, as
derived from Fig. 2 of Ref.~13. 
The island size for the lowest CO partial pressure
(the `clean' case) is only a lower bound for $L_{1/2}$, since the first
monolayer coalesces prior to second layer nucleation. 
The full symbols
show the variation of the additional step edge barrier $\Delta
E_S$ with $p_{\rm CO}$. The data points represented by full circles are
determined by use of equation (\ref{method2}). 
Due to the applied CO partial pressure, the
additional step edge barrier increases dramatically from 0.12 eV to
0.36 eV. For the calculation, $\alpha$ was adjusted between
$1/18$ and $1/15$ to account for the island
geometry (compare to Fig. 1 of Ref.~13).

The results are compared to the step
edge barrier obtained from the TDT theory \cite{tersoff}. 
We have adapted their expression for the nucleation rate on a 
circular island (Eq.(3) of Ref.~5) 
to the notation
of the present paper, which yields
\begin{equation}
\label{nuprimeTDT}
1/\nu'= \left(\frac{12 \ln 2}{\gamma F N \nu \alpha^4 L_{1/2}^6} -
\frac{52}{49} \frac{\alpha^2 L_{1/2}^2}{\nu^2} \right)^{1/2} - 
\frac{12}{7} \frac{\alpha L_{1/2}}{\nu}.
\end{equation}
A reasonable choice\cite{Venables} for the capture
number in this case is $\gamma = 3$. The resulting data
is shown in Fig. \ref{Fig-CO} as full triangles. It is obvious that the TDT
approach seriously underestimates the magnitude of the step edge barrier for
large barriers; for example, for 
$p_{\rm CO} = 1.85 \times 10^{-9}$ mbar the TDT formula 
yields\cite{Note:Pref} 
$\Delta E_S = 0.28$ eV instead of 0.36 eV. 
We note that the strong barrier condition $\nu/\nu' >  L$ is valid
only for the CO partial pressures in
excess of $1 \times 10^{-10}$ mbar, hence the barrier obtained using
the LAM can be trusted only for these pressures. Attempts to improve
the estimates for small barriers using the upper bound (\ref{bound})
on the nucleation rate result in negative values for
$\nu'$, which shows that the bound is too rough to be useful.

We still need to address the consistency of our assumption that
$i^\ast = 1$. In fact, at 400 K a Pt-dimer may not be
considered to be a stable nucleus \cite{Bott}, because Pt-dimer
breaking energies on Pt(111) obtained in experiment\cite{markov,Ehrlich99} 
and calculation\cite{Boisvert}
vary from 0.49 eV to 0.81 eV. It is
likely
that at 400 K a trimer is a stable nucleus. At this temperature, the
condition $\tau_{\rm tr} \ll \tau_{\rm dis} \ll \Delta t \cdot
(\tau_{\rm tr}/\tau)$
for the scaling regime III of Section \ref{I2}
is fulfilled. For example, for the largest
CO partial
pressure and assuming a dimer breaking energy of 0.65 eV one obtains
with the other
values above $\tau_{\rm tr} = 1.4 \times 10^{-6}$ s,
$\tau_{\rm dis} = 1.9 \times 10^{-5}$ s and
$\Delta t \cdot (\tau_{\rm tr}/\tau) = 2.4 \times 10^{-3}$ s. 
Unfortunately the nucleation rate in regime III
depends explicitly on the (unknown) dimer dissociation time, and
therefore the expression (\ref{LcIII}) cannot be used for a quantitative
analysis. Since nucleation on a top terrace is more difficult for
$i^\ast > 1$  
compared to $i^*=1$, and at a given island size
a more efficient step edge barrier is necessary for 
nucleation to occur, the step edge barrier
obtained under the assumption $i^\ast = 1$
is only a lower bound. 

We conclude, therefore,
that the step edge barriers presented in Fig. \ref{Fig-CO} for pressures
$p_{\rm CO} > 10^{-10}$ mbar are 
lower bounds on the true values. For smaller pressures this need no longer
be true, because our expression (\ref{omega}) underestimates the
nucleation rate. For the data point corresponding to `clean' growth
($p_{\rm CO} = 5 \times 10^{-12}$ mbar) the situation is further complicated
by the fact that the measurement yields only a lower bound on $L_{1/2}$.
Therefore at this point a direct comparison with other measurements 
\cite{kyuno98} and calculations \cite{feibelman98} of the additional barrier
for clean steps on Pt(111) is not possible.

\section{The size of the top terrace of a wedding cake}
\label{Wedding}

In this section we apply the LAM
to the growth of multilayer films in the ``wedding cake''
regime \cite{jk97,politi97,kalff99}. Wedding cakes are pyramidal
mounds of fixed lateral size which grow on the template of
the first layer islands. They constitute the typical growth
morphology under conditions of strong step edge barriers
in the sense of Section \ref{Nucleation}. In the limit of
infinite step edge barriers ($\ell_{\rm ES} \to \infty$) 
the wedding cakes have
a characteristic pointed shape which can be deduced \cite{jk97}
from the observation \cite{cohen} that the coverages of the
exposed layers follow a Poisson distribution. When 
$\ell_{\rm ES}$ is finite but large the overall shape remains
unchanged, however the pointed tip is replaced by a 
flat top terrace, whose typical size $L_{\rm top}$ directly reflects
the strength of the step edge barriers \cite{politi97,kallabis}. 

For an order of magnitude estimate it suffices to note that
the approximate reproduction of the mound shape after the growth
of an additional layer requires on average one nucleation
event per top terrace and deposited monolayer, and therefore
\cite{physbl}
\be
\label{Ltop}
\omega (L_{\rm top})/ F \approx 1.
\ee
The purpose of this section is to refine this argument and
to develop a simple analytic model for the size distribution
of top terraces. A detailed comparison with growth experiments
on Pt(111) \cite{kalff99} will be carried out in Section 
\ref{Pt111}, while extensions and modifications of the theory
are discussed in Sections \ref{1D} and \ref{Variants}. 

\subsection{Model for the dynamics of the top terrace}
\label{Model}

Any degree of interlayer transport couples the growth of
all atomic layers \cite{cohen}. However, 
the observation \cite{politi97,kalff99} that for strong
barriers the shape is affected only near the top of 
the wedding cake suggests that the main features can
be captured in a model which decouples the growth of the
top terrace from the rest of the structure.   
This is achieved most simply by assuming
that the top terrace grows on a {\em base terrace} of fixed size
$\Lambda$, which is bounded by a step with an infinite edge barrier. 
Then all atoms landing on the base terrace or on the top terrace contribute
to the growth of the latter, and its size evolves according to 
\begin{equation}
\label{L(t)}
L(t) = \sqrt{Ft} \Lambda = \sqrt{\theta} \Lambda,
\end{equation} 
where time $t$ and coverage $\theta = F t$ are counted from the event
of nucleation. The interpretation of $\Lambda$ as the size of the base
terrace should not be taken too literally, since we will allow 
for $L$ to become larger than $\Lambda$; rather, 
$\Lambda$ sets the scale of the
top terraces and will eventually be fixed self consistently. 
In principle the constraint $L \leq \Lambda$ could be built
into the growth law for $L(t)$, but we have found that this does
not improve the agreement with the experimental data to be
presented in Section \ref{Pt111}. An example of such a modified
growth law will be briefly discussed in Section \ref{Variants}.

The growth law (\ref{L(t)}) holds until a new top terrace nucleates on
the previous one, at which point $L$ is reset to 0.
The growth is assumed to be {\em deterministic},
which implies that
fluctuations in the deposition and diffusion events on 
the terrace are ignored. At least for sufficiently large
terraces these should be negligible compared to the fluctuations
in the times between subsequent nucleation events, which are
taken into account explicitly.  
The probability
for no nucleation to occur up to time $t$ (coverage $\theta = Ft$) is
\begin{equation}
\label{P0}
P_0(\theta) = \exp[-\int_0^t dt' \; \omega(L(t'))] = 
\exp \left[ - \frac{2}{k+2} \Omega \Lambda^k \theta^{(k+2)/2}
\right].  
\end{equation}
Here the general expression (\ref{omegagen}) 
for the nucleation rate as a function 
of island size and the relation (\ref{L(t)}) for the time dependence of $L$
have been used.
The distribution $P(\theta)$ of coverages between
subsequent nucleation events is then given by
$P(\theta) = - dP_0/d \theta$.
To fix the value of $\Lambda$ we observe that, in order
for the mound morphology to be reproduced after the growth of one layer,
the mean of $P(\theta)$ should equal unity. This implies the condition
\begin{equation}
\label{thetamean}
\int_0^\infty d\theta \; \theta P(\theta) = \int_0^\infty P_0(\theta) = 1.
\end{equation}
Inserting (\ref{P0}) we find
\begin{equation}
\label{Lambda}
\Lambda = \left( \frac{k+2}{2} C_k \right)^{1/k} \Omega^{-1/k}
\end{equation}
where
\be
\label{Ck}
C_k = [\Gamma((k+4)/(k+2))]^{(k+2)/2}
\ee
and $\Gamma$ denotes the $\Gamma$-function. 
For example, in the case $i^\ast = 1$ we have $k=5$ and 
$\Omega = \alpha^3 F/\nu'$ and (\ref{Lambda}) yields
\be
\label{Lambda1}
\Lambda = 1.193 \cdot \alpha^{-3/5} (\nu'/F)^{1/5}.
\ee 

It is plausible (and can be checked by a formal argument) that the
probability $Q(L)$ of finding a top terrace of size $L$ is proportional to the
probability that no nucleation has occured up to coverage $\theta = 
(L/\Lambda)^2$.
Using the expression (\ref{P0}) and transforming variables from $\theta$
to $L$ using (\ref{L(t)}) we find
\begin{equation}
\label{Q}
Q(L) = (d \theta/dL) \langle \theta \rangle^{-1} P_0(L^2/\Lambda^2) =
\frac{2L}{\Lambda^2} \exp[-C_k (L/\Lambda)^{k+2}].
\end{equation}
Here the condition $\langle \theta \rangle = 1$ has been enforced 
through the relation (\ref{Lambda}).

The distribution increases linearly for small $L$ and
is rather abruptly cut off at $L \approx \Lambda$ (see Fig. \ref{Fig-Q}). 
Its moments can be expressed
in terms of $\Gamma$-functions. In particular, the mean size
of the top terrace is given by
\begin{equation}
\label{meanr}
\langle L \rangle / \Lambda = \frac{2}{3} \frac{\Gamma((k+5)/k+2))}{\Gamma(
(k+4)/(k+2))^{3/2}}.
\end{equation}
For $k \to \infty$ the factor on the right hand side approaches
2/3, which corresponds to a linear distribution $Q(L) = 2L/\Lambda^2$
with a sharp cutoff at $L = \Lambda$. Already for $k=5$ the 
ratio $\langle L \rangle / \Lambda \approx 0.692$ is very
close to 2/3. 
Another consequence of the skewed shape of $Q(L)$ is that the most
probable value $L_{\rm max}$ is considerably larger than the mean,
and is generally given by  
\begin{equation}
\label{Lmax}
L_{\rm max}/\Lambda = (k+2)^{-1/(k+2)} \Gamma((k+4)/(k+2))^{-1/2}.
\end{equation}
For $k = 5$, $L_{\rm max}/\Lambda \approx 0.80$, and 
$L_{\rm max}/\Lambda \to 1$ for $k \to \infty$.

\subsection{Multilayer growth on Pt(111)}
\label{Pt111}

The model for the dynamics of the top terrace is substantiated by its
application to mound growth on Pt(111). As shown in Fig. 
\ref{Fig-mounds}, after
deposition of 37.1 ML Pt with a deposition rate $F = 1.3 \times
10^{-2}$ ML/s in a partial pressure $p_{\rm CO} = 1.9 \times 10^{-9}$ mbar
of carbon monoxide at 440 K mounds form, which are built from terraces
scattering slightly around the shape of an equal-sided hexagon. The
differentiated representation (morphology appears illuminated from the
left) of the scanning tunneling topographs allows to distinguish all atomic
steps, in particular the steps bounding the top terrace and its base terrace.
The applied CO partial pressure during growth results in a significant decoration of
step edges by CO molecules, while the equilibrium coverage of the terraces is
negligible at 440 K (about $7 \times 10^{-3}$ ML \cite{Poelsema}). As already
mentioned in Section \ref{Pt(111)islands}, the step decoration with CO gives rise
to a significant increase in the additional step edge barrier \cite{Kalff98}.

The
distributions of the sizes $L$ of the top terraces and $\Lambda$
of the base terraces determined for about 150
mounds are presented in Figure \ref{Fig-Q}.
The mean base terrace size in units of
edge sites is $\langle \Lambda \rangle = 212$. The relatively narrow
distribution of base terraces
shown in Figure \ref{Fig-Q} supports our model assumption of
a constant base terrace size. The mean top terrace size
in units of edge sites is $\langle L \rangle = 142$. The ratio of the
two
averages $\langle L \rangle / \langle \Lambda \rangle = 0.67$. The most
probable value $L_{\rm max}$ is considerably larger than its mean and is
found in the size
interval $ [0.78,0.81] \langle \Lambda \rangle$.  

For comparison with
the growth model
developed in the previous section a choice for the size of $i^*$ is
necessary.
As discussed in section \ref{Pt(111)islands}, the scaling regime III
with $i^*=2$
most probably applies here. Consequently we perform the comparison 
of the shape of the size distribution for
$i^*=2$, i.e. $k=6$ in (\ref{Q}); note however that 
(\ref{Q}) depends rather weakly on $k$, and therefore
the results assuming $i^*=1$, $k=5$ are almost identical. 
The experimental ratio
$\langle L \rangle / \langle \Lambda \rangle = 0.67$ is in excellent
agreement with the model prediction of $\langle
L \rangle / \Lambda = 0.687$, and the range of 
experimental values for
$L_{\rm max}$ agrees very well with
the model prediction $L_{\rm max} = 0.81 \Lambda$.
The figure also shows the analytic expression (\ref{Q})
for the distribution of top terrace sizes.
Given the limited number of evaluated
mounds, the agreement is quite satisfactory.

Due to the uncertainty in the dimer bond strength, 
for the quantitative estimate of $\nu'$ and $\Delta E_S$ 
we will assume $i^*=1$; as discussed in Section
\ref{Pt(111)islands}, this
will provide a lower bound on the step edge barrier.
Rearrangement of formula
(\ref{Lambda1}) yields
\begin{equation}
\nu' = 0.414 \cdot \alpha^3 F \Lambda^5.
\end{equation}
Inserting the experimental deposition rate, choosing $\alpha =
1/12$ for equal-sided hexagons and a triangular lattice and
$\langle \Lambda \rangle =  212$ as determined experimentally, we
obtain for the hopping rate over step edges $\nu' = 1.32 \times
10^{6} {\rm s}^{-1}$. With the same assumptions as in section
\ref{Pt(111)islands} this translates into $\Delta E_S = 0.33$ eV,
in reasonable agreement
with the step edge barrier obtained at 400 K and the same
CO partial pressure in the analysis of second layer nucleation in 
Section \ref{Pt(111)islands}. The agreement indicates
that in both cases the CO concentration at the step edges was close
to saturation, and supports the
consistency of our two approaches to the determination of the additional
step edge barrier from submonolayer islands and multilayer mounds.

\subsection{The one-dimensional case}
\label{1D}

The considerations of Sections 
\ref{Nucleation} and \ref{Model} are easily extended
to one-dimensional islands. For an island of length $\ell$
the mean interval time between subsequent deposition events is
$\Delta t =1/F\ell$, while the residence time is 
$\tau = \ell /2 \nu'$. 
Therefore the nucleation rate for strong step edge barriers
and $i^\ast = 1$ is
\be
\omega = {\tau\over (\Delta t)^2} = {F^2 \ell^3 \over 2 \nu'}.
\ee
The time evolution for the size of the top terrace is now
$\ell(t)=Ft \lambda =\theta \lambda$, 
with $\lambda$ denoting the length of the base terrace,
and the probability $P_0(\theta$) that no
nucleation event has taken place after a coverage $\theta$ is 
\be
P_0(\theta) = \exp[-(F \lambda^3 /8 \nu') \; \theta^4].
\ee
As before, $\lambda$ is determined
by the condition (\ref{thetamean}). This implies
\be
\label{Lambda1D}
\lambda = (\Gamma(1/4)^{4/3}/2^{5/3}) (\nu'/F)^{1/3} 
\approx 1.75 \; (\nu'/F)^{1/3}.
\ee
To summarize, for $i^\ast = 1$ the size of the top terrace
scales with the Ehrlich-Schwoebel length as
$\ell_{\rm ES}^{-1/3}$ in $1+1$ dimensions, and as
$\ell_{\rm ES}^{-1/5}$ in $2+1$ dimensions.
This agrees with the results found in Refs.~15 and 25
through a less quantitative method. 

\subsection{Deterministic nucleation and other modifications}
\label{Variants}

To assess the importance of the fluctuations in the nucleation
times, we now consider a model where nucleation occurs at
fixed time intervals corresponding precisely to the growth
of one monolayer. The base terrace size $\Lambda$ is determined
through the requirement that these nucleation times should correspond
to the {\em maximum} of the distribution, i.e. rather
than fixing the mean of $P(\theta)$ we fix the position of its peak. 
Noting that $P(\theta) = - d P_0/d \theta = F^{-1} \omega P_0$,
the condition $dP/d \theta = 0$ for the maximum can be written as
\be
\label{Pmax}
F \frac{d \omega}{d \theta} = \omega^2,
\ee
where the nucleation rate (\ref{omegagen}) has been expressed in terms
of the coverage using (\ref{L(t)}). Solving (\ref{Pmax}) and setting
the corresponding coverage to unity yields
\be
\label{Lambda2}
\Lambda = (k/2)^{1/k} \Omega^{-1/k}.
\ee
The numerical coefficients in (\ref{Lambda}) and (\ref{Lambda2}) coincide
within a few percent for $5 \leq k \leq 8$. Thus for most purposes
the simpler relation (\ref{Lambda2}) can be used. Similarly for the
one-dimensional case with $i^\ast = 1$ the deterministic approach yields
\be
\label{1Ddet}
\Lambda = 6^{1/3} (\nu'/F)^{1/3} 
\approx 1.82 \; (\nu'/F)^{1/3},
\ee
very close to (\ref{Lambda1D}). 

An advantage of the deterministic approach is that more complicated
growth laws for the top terrace size $L(t)$ are easily incorporated.
Consider for example a situation where, instead of fixing the
size of the base terrace, the base terrace is allowed
to grow such that the area $A_0$ of its {\em uncovered} part remains
constant. For consistency we should then also allow some fraction
$1 - \mu$ of the mass deposited onto the top and the base terraces
to contribute to the growth of the latter. This implies that
the area $A(\theta)$ of the top terrace evolves 
according to $dA/d\theta = \mu(A + A_0)$, which yields the growth law
\be
\label{rim}
A(\theta) = A_0(e^{\mu \theta} -1).
\ee
The additional parameter $\mu$ is fixed by the requirement that
nucleation should occur at $\theta = 1$, at which point the top terrace
becomes the base terrace and hence $A(1) = A_0$ or $\mu = \ln 2$.
Using the expression (\ref{omegagen}) for the nucleation rate and
imposing the relation (\ref{Pmax}) at $\theta = 1$ then fixes $A_0$
to be
$
A_0 = \alpha (k \mu)^{2/k}
\Omega^{-2/k}
$,
and the size of the base terrace at the time of nucleation is 
\be
\label{L0}
L_0 = \sqrt{A_0/\alpha} = (k \ln2)^{1/k} \Omega^{-1/k},
\ee
which differs from (\ref{Lambda2})
by less than 10 $\%$ for $5 \leq k \leq 8$.

\section{Conclusions}

The main purpose of this paper has been to establish some
{\em quantitatively}
accurate relations between microscopic parameters and large
scale morphology for homoepitaxial growth 
in the presence of destabilizing step edge barriers,
and to demonstrate their usefulness for the well-studied case
of Pt/Pt(111). Perhaps surprisingly, we found that a quantitative
analysis of nucleation is simpler with strong step edge barriers
than without them: Due to the homogeneity of the adatom occupation 
probability, and the relation $p_{\rm nuc} = p_2$, our expression
(\ref{omega}) for the nucleation rate is exact in the strong barrier
regime for any island shape. It would be most desirable to obtain
exact results also for intermediate and weak barriers as well as
for reversible aggregation, but for the reasons described in Sections 
\ref{Weak} and \ref{I2} such an extension is far from straightforward.

On a conceptual level, the reasons for the failure of the
rate equation ansatz (\ref{rateeq})
in the present problem need to be better understood. 
We have identified one regime in the case $i^\ast = 2$
in which our analysis agrees with the rate equation theory \cite{tersoff}
(see Section \ref{I2});
such a regime exists\cite{maass99} also for  
$i^\ast > 2$, but the precise conditions for the
validity of (\ref{rateeq}) are not clear. In this context
it is worth pointing out that the standard rate equation approach also fails
in one-dimensional nucleation \cite{pimp92,villain92,kallabis98}. 
This suggests that the rate equation treatment may generally be expected
to be problematic in situations involving low-dimensional or confined 
geometries.

\vspace*{0.5cm}

\noindent
{\bf Acknowledgements.} J.K. wishes to thank Pavel \v{S}milauer and 
Kelly Roos for useful discussions. 
T.M. acknowledges experimental help from Matthias Kalff and
inspiration by the ideas and simulations of Michael Hohage.
Support by DFG through SFB237 (J.K.)
and through a Heisenberg fellowship (T. M.), by the 
Alexander von Humboldt foundation (P.P.) 
and by Volkswagenstiftung (J.K.) is gratefully acknowledged.   

\appendix

\section{Calculation of $p_{n+1}$}
\label{pn+1}

Here we compute the probability $p_{n+1}$ of finding $n+1$ 
(noninteracting) adatoms simultaneously on the island. 
Denote by $t_i$ the time interval between the arrivals of atoms
$i$ and $i+1$, and by $\tau_i$ the residence time of atom $i$,
with $i = 1,...,n$. The first atom arrives at time $t=0$, the
second at time $t = t_1$ and so on. For all $n$ atoms to still
be on the island when the $n+1$'st atom arrives, the residence
times have to satisfy the conditions $\tau_i > t_i + t_{i+1} + 
... + t_n$. The $t_i$ and $\tau_i$ are 
exponential random variables with mean $\Delta t$ and $\tau$,
respectively. Thus
$$
p_{n+1} = (\tau \; \Delta t)^{-n} 
\prod_{i=1}^n \int_0^\infty dt_i \; e^{-t_i/\Delta t} 
\prod_{i=1}^n \int_{t_i + t_{i+1} + ... + t_n}^\infty 
d\tau_i \; e^{-\tau_i/\tau} = 
$$
$$
= (\Delta t)^{-n} 
\prod_{i=1}^n \int_0^\infty dt_i \; e^{-t_i/\Delta t} 
\prod_{i=1}^n e^{-(t_i + t_{i+1} + ... + t_n)/\tau} = 
(\Delta t)^{-n} 
\prod_{i=1}^n \int_0^\infty dt_i \; \exp[-t_i(1/\Delta t + i/\tau)] =
$$
\be
\label{pn1}
= \prod_{i=1}^n (1 + i \Delta t/\tau)^{-1}.
\ee
For $\tau \ll \Delta t$ this reduces to 
\be
\label{pn2}
p_{n+1} \approx \frac{1}{n !} \left(\frac{\tau}{\Delta t} \right)^n.
\ee

\section{Critical island size for $i^\ast = 2$}
\label{LcI2}

In the different regimes described in Section \ref{I2},
the critical island size for second layer nucleation is given by
the expressions
\be
\label{LcI}
L_c \sim \left(\frac{\nu'}{F} \right)^{2/7} \left( 
\frac{\nu^2 \tau_{\rm dis}}{\nu'} \right)^{-1/7} N^{-1/7}
\;\;\;\;\;\;
({\rm regime \;\; I})
\ee
\be
\label{LcII}
L_c \sim \left(\frac{\nu'}{F} \right)^{1/5}  N^{-1/10}
\;\;\;\;\;\;
({\rm regime \;\; II})
\ee
\be
\label{LcIII}
L_c \sim \left(\frac{\nu'}{F} \right)^{1/4} 
(\nu \tau_{\rm dis})^{-1/8} N^{-1/8},
\;\;\;\;\;\;
({\rm regime \;\; III})
\ee 
and for general $i^\ast$ the nucleation rate (\ref{caseIIgen}) for
regime II yields 
\be
\label{Lcgen}
L_c = \left(\frac{(4 + 3i^\ast) i^\ast !}{2 \alpha^{2 i^\ast + 2}}
\right)^{1/(4 + 3 i^\ast)} 
\left(\frac{\nu'}{F} \right)^{i^\ast/(4 + 3 i^\ast)} 
N^{-1/(4 + 3 i^\ast)}.
\ee

\begin{figure}
\caption{Critical island size $L_{1/2}$ (open circles) and the resulting
additional step edge barrer $\Delta E_S$
(full symbols) as a function of CO partial
pressure. Full circles were calculated according to the LAM formula
(\ref{method2}), while full triangles were obtained using the TDT
formula (\ref{nuprimeTDT}).}
\label{Fig-CO}
\end{figure}

\begin{figure}
\caption{Size distribution of top terraces. The full line shows a
histogram obtained from the evaluation of 145 mounds, and the thick dotted
line shows the (appropriately normalized) theoretical prediction 
(\ref{Q}). The thin dashed line is included to illustrate the narrowness
of the distribution of base terraces.}
\label{Fig-Q}
\end{figure}

\begin{figure}
\caption{Typical STM image of mounds appearing on the surface after
the deposition of 37.1 ML Pt/Pt(111) in a partial CO pressure of 
$1.9 \times 10^{-9}$ mbar. Eight top terraces and the corresponding
base terraces can be seen.}
\label{Fig-mounds}
\end{figure}


\begin{thebibliography}{99}
\bibitem{Note:Pref}
In Ref.~13 a smaller value was obtained because
of an incorrect treatment of the prefactor in (\ref{nuprimeTDT}).
\end{thebibliography}
 
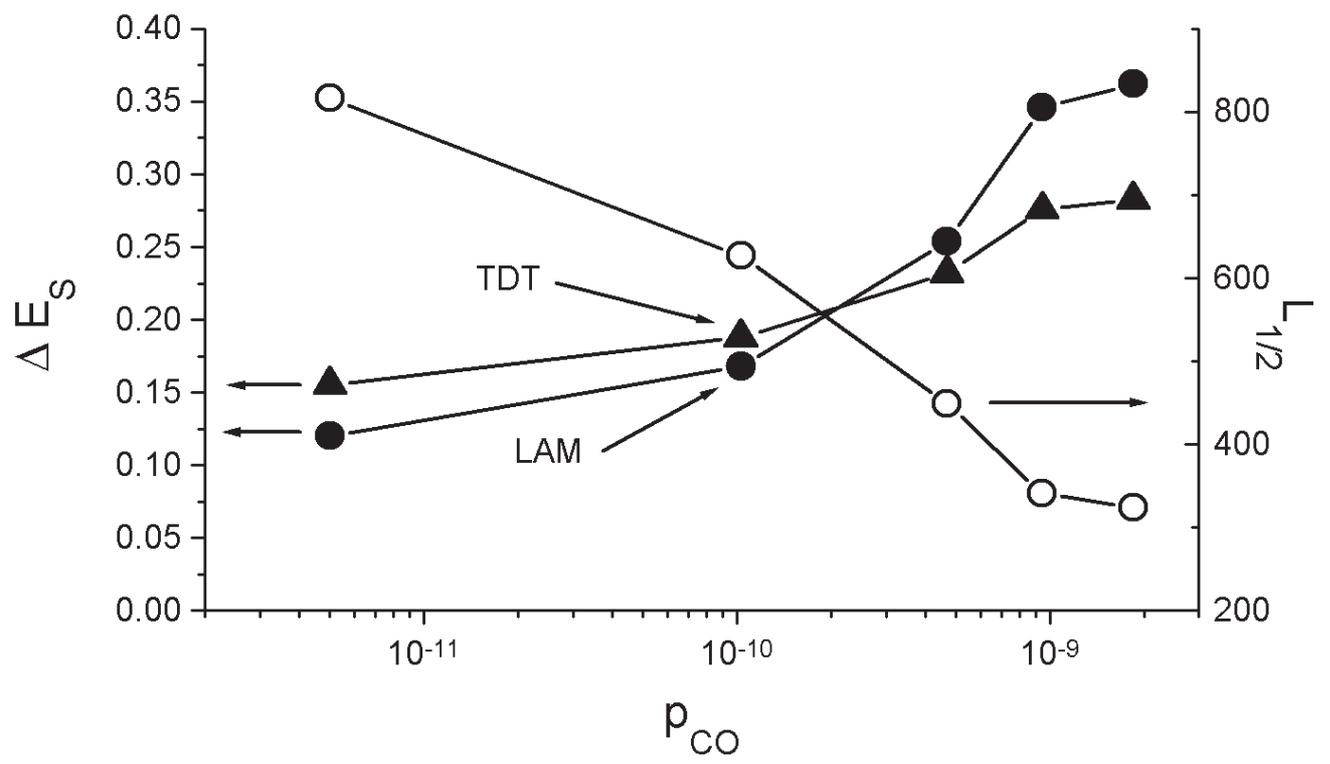
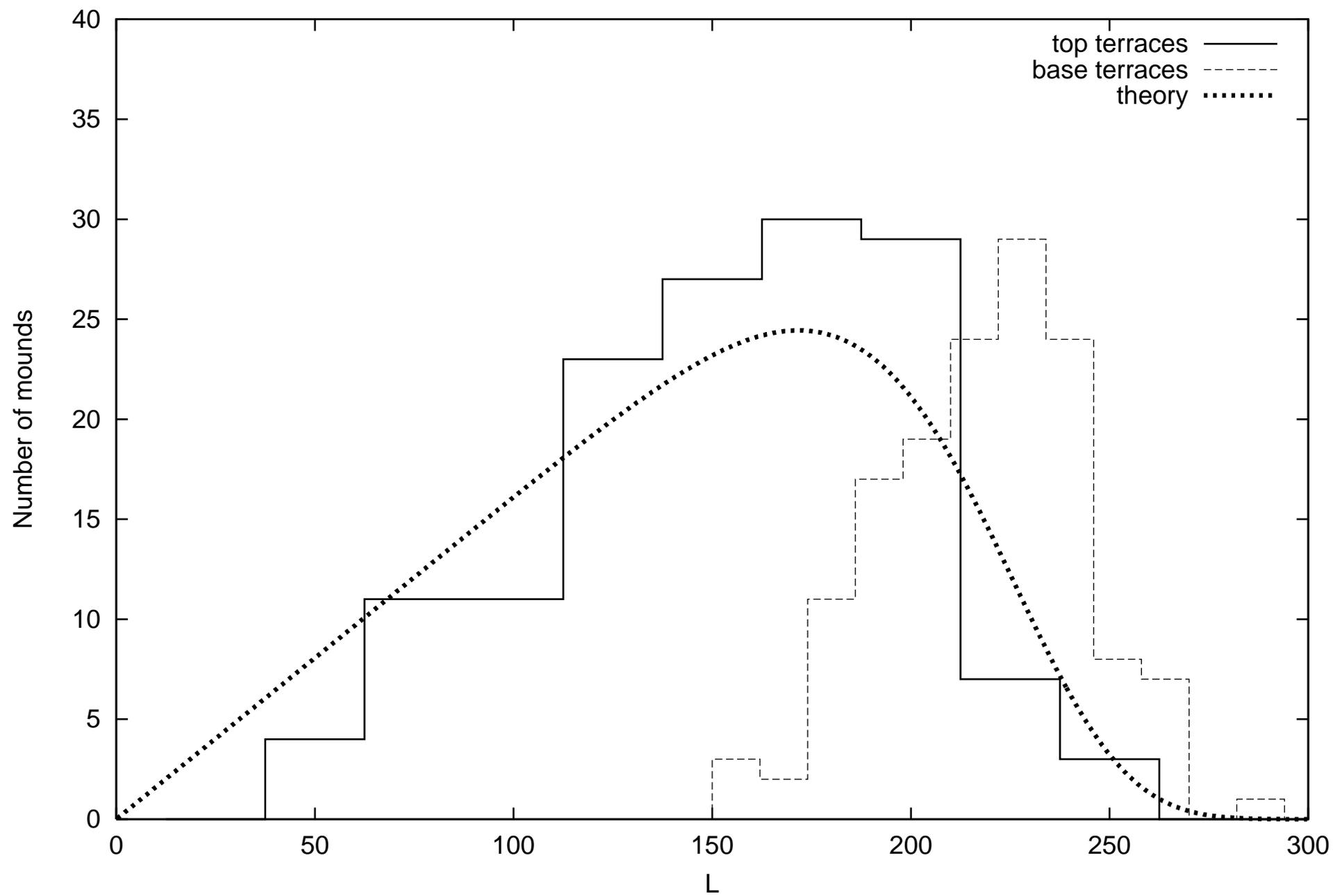
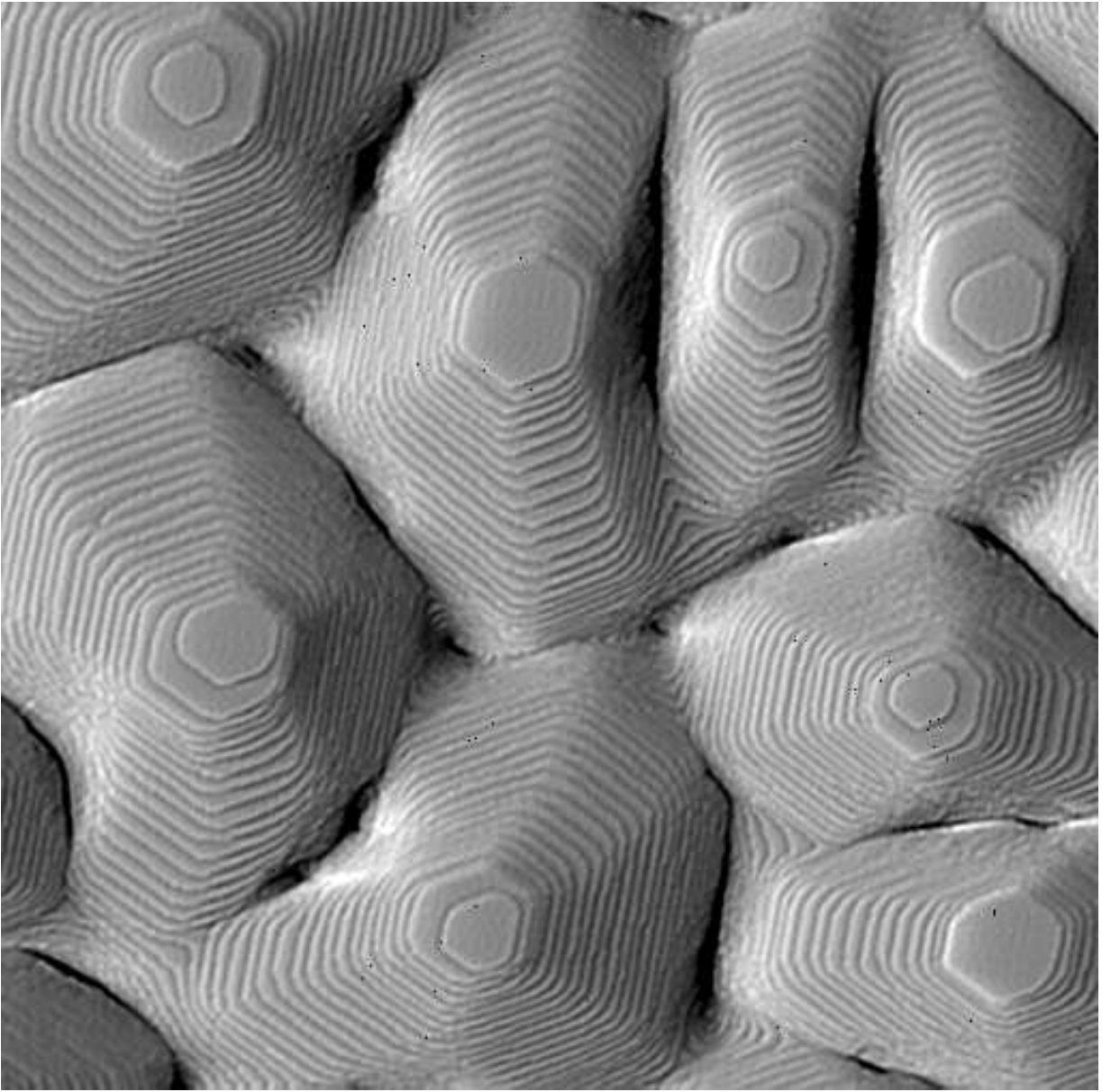
\end{document}